\documentclass[11pt,a4paper]{article}
\usepackage{times,latexsym}
\usepackage{url}
\usepackage[T1]{fontenc}

\usepackage[acceptedWithA]{tacl2021v1}

\usepackage{xspace,mfirstuc,tabulary}

\newif\iftaclinstructions
\taclinstructionsfalse 
\iftaclinstructions
\renewcommand{\confidential}{}
\renewcommand{\anonsubtext}{(No author info supplied here, for consistency with
TACL-submission anonymization requirements)}
\newcommand{\instr}
\fi

\iftaclpubformat 

\else

\fi

\usepackage{amsmath,amsfonts,bm,bbm}
\usepackage{mathrsfs,amssymb,mathtools}
\usepackage{amsthm}

\def\eqref#1{equation~\ref{#1}}

\def\1{\bm{1}}

\DeclareMathAlphabet{\mathsfit}{\encodingdefault}{\sfdefault}{m}{sl}
\SetMathAlphabet{\mathsfit}{bold}{\encodingdefault}{\sfdefault}{bx}{n}

\usepackage{microtype}

\usepackage{inconsolata}

\usepackage{graphicx}

\usepackage[utf8]{inputenc} 
\usepackage[T1]{fontenc}    
\usepackage{hyperref}       
\usepackage{url}            
\usepackage{booktabs}       
\usepackage{amsfonts}       
\usepackage{nicefrac}       
\usepackage{microtype}      
\usepackage{xcolor}         

\usepackage{color, colortbl}
\usepackage{paralist}
\usepackage{enumitem}
\usepackage{multirow}
\usepackage{caption}

\usepackage{graphicx}
\usepackage{subcaption}
\usepackage{algorithm}
\usepackage{algorithmic}
\usepackage{makecell}
\usepackage{tikz}
\usepackage{pgfplots}
\usepackage{adjustbox}
\usepackage{wrapfig}
\usepackage{soul}
\usepackage{comment}

\usepgfplotslibrary{fillbetween}
\usetikzlibrary{patterns}

\definecolor{darkgreen}{rgb}{0.0, 0.5, 0.0}
\definecolor{fuchsia}{rgb}{0.57, 0.36, 0.51}
\definecolor{amber}{rgb}{1.0, 0.49, 0.0}
\definecolor{apricot}{rgb}{0.98, 0.81, 0.69}
\definecolor{auro}{rgb}{0.43, 0.5, 0.5}

\definecolor{darkpink}{rgb}{0.847, 0.105, 0.376}
\definecolor{bluevar}{rgb}{0.117, 0.533, 0.89}
\definecolor{yellowvar}{rgb}{1, 0.75,0.02}
\definecolor{darkgreenvar}{rgb}{0.0, 0.3, 0.250}
\definecolor{lightgreenvar}{rgb}{0.0, 0.3, 0.250}
\definecolor{chromevar}{rgb}{0.905,0.43,0.31}
\definecolor{orangevar}{rgb}{0.96,0.49,0}
\definecolor{weirdyellowvar}{rgb}{1,0.83,0.039}
\definecolor{weirdgreenvar}{rgb}{0.50,0.92,0.6}
\definecolor{darkbrownvar}{HTML}{540B0E}
\definecolor{lightbrownvar}{HTML}{9E2A2B}

\definecolor{crazybluevar}{HTML}{03045e}

\definecolor{lightbluevar}{HTML}{6096ba}
\definecolor{darkbluevar}{HTML}{274c77}

\definecolor{lightvioletvar}{HTML}{f4acb7}
\definecolor{darkvioletvar}{HTML}{907ad6}

\definecolor{sapgreenvar}{HTML}{80ED99}
\definecolor{sapgreenvar2}{HTML}{57CC99}

\usetikzlibrary{positioning}

\newcommand{\revision}[1]{#1}
\newcommand{\revisionb}[1]{#1}

\definecolor{darkgreen}{rgb}{0.0, 0.5, 0.0}
\definecolor{darkblue}{rgb}{0.0, 0.0, 0.5}
\definecolor{darkred}{rgb}{0.5, 0.0, 0.0}

\makeatletter
\newcommand{\oset}[3][0ex]{\mathrel{\mathop{#3}\limits^{\vbox to#1{\kern-2\ex@\hbox{$\scriptstyle#2$}\vss}}}}
\makeatother

\usepackage{letltxmacro}
\LetLtxMacro\orgvdots\vdots
\LetLtxMacro\orgddots\ddots
\usepackage{amssymb}
\usepackage{pifont}

\usepackage[colorinlistoftodos,prependcaption,textsize=small]{todonotes}

\usepackage{amsthm}
\theoremstyle{definition}
\newtheorem{definition}{Definition}[section]

\usepackage[symbol]{footmisc}

\def\bp{{\mathbf{p}}}
\def\bx{{\mathbf{x}}}

\title{Towards More Realistic Extraction Attacks: An Adversarial Perspective}

\author{Yash More\Thanks{Equal contribution}\\
  McGill University, Mila \\
  \texttt{yash.more@mila.quebec} \\\And
  Prakhar Ganesh$^*$ \\
  McGill University, Mila \\
  \texttt{prakhar.ganesh@mila.quebec} \\\AND
  Golnoosh Farnadi \\
  McGill University, Mila \\
  \texttt{farnadig@mila.quebec} \\}

\date{}

\begin{document}
\maketitle
\begin{abstract}
Language models are prone to memorizing their training data, making them vulnerable to extraction attacks. While existing research often examines isolated setups, such as a single model or a fixed prompt, real-world adversaries have a considerably larger attack surface due to access to models across various sizes and checkpoints, and repeated prompting. In this paper, we revisit extraction attacks from an adversarial perspective---with multi-faceted access to the underlying data. We find significant churn in extraction trends, i.e., even unintuitive changes to the prompt, or targeting smaller models and earlier checkpoints, can extract distinct information. By combining multiple attacks, our adversary doubles ($2 \times$) the extraction risks, persisting even under mitigation strategies like data deduplication. We conclude with four case studies, including detecting pre-training data, copyright violations, extracting personally identifiable information, and attacking closed-source models, showing how our more realistic adversary can outperform existing adversaries in the literature.


\end{abstract}

\section{Introduction}
\label{sec:introduction}

Large language models (LLMs) have grown considerably in size~\citep{MetaAI_Llama3,zhao2023survey}, and have become integral to a wide range of tasks such as knowledge retrieval, question answering, code generation, and machine translation.

To complement this growing scale, LLMs are often trained on large amounts of data~\citep{penedo2024fineweb,cerebras2023slimpajama,gao2020pile,raffel2020exploring} that may include private information, especially if scraped from the web. As LLMs are prone to memorizing the data they have been trained on, they can be prompted to expose sensitive contexts---making it easier for an adversary to extract information. Naturally, a question arises: how big is the risk imposed due to \textit{memorization}? 

\textit{Extraction attacks} offer an empirical framework to quantify the information leakage in the presence of an adversary. \revisionb{A commonly studied extraction attack is discoverable memorization~\citep{carlini2023quantifying,kassem2024alpaca}, where the adversary extracts targeted information from the training data by prompting the model with a portion of a sentence from the training data to extract the rest. Discoverable memorization has been used in many adversarial settings, including membership inference~\citep{maini2024llm}, data contamination detection~\citep{ravaut2024much}, and copyright violations~\citep{karamolegkou2023copyright}, among others.}




Current extraction attacks study memorization trends in LLMs across isolated settings like model sizes, generation hyperparameters, and learning dynamics \citep{carlini2023quantifying}. While effective, they underestimate the risk posed due to a multi-faceted access to the underlying data in the current LLM ecosystem. For instance, we show that an adversary can exploit the sensitivity of LLMs to prompt structure, length, and content, to amplify the information gained. The current accessibility to frequently updated model sizes~\citep{MetaAI_Llama3}; checkpoints \citep{biderman2023pythia, groeneveld-etal-2024-olmo} and a large array of model families such as Llama~\citep{MetaAI_Llama3}, Gemini~\citep {team2023gemini}, and Falcon~\citep {almazrouei2023falcon}, can also create higher extraction risks.

In this paper, we study a more realistic scenario and explore the actual risks posed by composite extraction attacks that can combine information from multiple attacks. More specifically, we ask:

\begin{enumerate}[leftmargin=*, parsep=0pt, itemsep=0pt, topsep=0pt]
    \item \textbf{Can adversaries exploit prompt sensitivity and access to multiple checkpoints?} We find that extraction attacks are sensitive to the prompt design, extracting over $20\%$ more data with even minor, unintuitive changes (\S \ref{sec:churn_prompt}). Similarly, we find that an adversary with access to multiple model checkpoints can increase the extraction rates up to $1.5 \times$ (\S \ref{sec:churn_model}). Thus, an adversary with multi-faceted access can extract far more data than previously observed.
    \item \revision{\textbf{Should the adversary always attack the most vulnerable setup?} Vulnerable setups are also more expensive to attack. Interestingly, we find that on a limited adversarial budget, using composite attacks on less vulnerable but cheaper setups can cause more information leakage (\S\ref{sec:cost}).}
    \item \textbf{Is data deduplication effective against composite attacks?} We find that data deduplication does reduce memorization, as expected~\citep{carlini2023quantifying}. However, adversaries can still exploit the prompt structure and multiple checkpoints to extract more information (\S \ref{sec:realistic_deduped}), hence our concerns persist despite deduplication.
    \item \textbf{How are downstream applications affected by a stronger adversary?} 
    We performed four separate case studies and found that our more realistic adversary improves the $p$-value of dataset inference in open-source models by up to $2 \times$ (\S \ref{sec:dataset_inference}), the extraction of copyright violations by up to $20\%$ (\S \ref{sec:copyright}), the extraction rate of personally identifiable information (PIIs) by $1.5 \times$ (\S \ref{sec:pii}), \revision{and training data inference even in closed-source models by $16\%$ (\S \ref{sec:closed})}.
    
\end{enumerate}

\section{Background and Related Work}
\label{sec:related_work}

We first introduce relevant background on extraction attacks in LLMs, followed by an overview of related work on prompt sensitivity and training dynamics in LLMs. Finally, we describe the term \textit{churn} as it applies in our context.

\noindent\textbf{Extraction Attacks in LLMs.} 
Unintended memorization in LLMs can make it prone to information leakage~\citep{tirumala2022memorization, secretsharer_carlini,mattern2023membership,carlini2022membership}, particularly through extraction attacks~\citep{birch2023model,carlini2021extracting,carlini2023quantifying,nasr2023scalable}. 
Extraction attacks have gained significant attention in recent years, studied under two primary frameworks: \textit{Discoverable Memorization}~\citep{carlini2023quantifying,kassem2024alpaca,liu2023trustworthy,biderman2023emergent,tirumala2022memorization,huang2022large}, where the adversary attempts to extract targeted information, and \textit{Extractable Memorization}~\cite{nasr2023scalable,kandpal2022deduplicating,qi2024follow}, where the adversary attempts to extract any information about the training data.

We add to the growing body of research on targeted extraction attacks by highlighting the lack of a realistic adversary in the literature. We show the existence of a stronger adversary capable of combining information from various attacks, thus defining \textit{composite discoverable memorization} (\S \ref{sec:adversarial_strengths}). 

\noindent\textbf{Prompt Sensitivity in LLMs.}
LLMs are sensitive to changes in their prompts, leading to fluctuations in their performance~\citep{scalar.2310.11324,promptingmethods}. This sensitivity persists across varying model sizes and through fine-tuning and other downstream modifications~\citep{butterfly.2401.03729, zhu2023promptbench}.
The sensitivity of prompts can also be misused, and adversarial modifications to prompts can trigger the model to act in unintended ways~\cite{rossi2024early,liu2024jailbreaking,hubinger2024sleeper,liu2024jailbreaking}. 

While several overarching trends studying the impact of prompt design on extraction attacks are present in the literature~\citep{carlini2023quantifying,kassem2024alpaca,qi2024follow,tirumala2022memorization}, these trends are often studied in isolation. Motivated by the composability of privacy leakage~\citep{mcsherry2009privacy}, we argue that an adversary capable of repeated prompting can combine these trends, and extract more information about the training data than previously reported (\S \ref{sec:churn_prompt}).

\noindent\textbf{Training Dynamics of LLMs.} Several recent works have studied the training dynamics of LLMs over time~\citep{tirumala2022memorization,whatdoesrobertaknow,xia2023training}. In the context of memorization, \citet{biderman2023emergent} explored the impact of model size and intermediate checkpoints on the dynamics of memorization, revealing a considerable variance in memorized data over time and size. The practice of releasing models in various sizes and regularly updating them can thus increase the attack surface.
Similar to prompt sensitivity, we study how adversaries can also exploit access to multiple checkpoints to extract more data (\S \ref{sec:churn_model}).



\begin{figure*}
    \centering
    \includegraphics[width=0.9\textwidth]{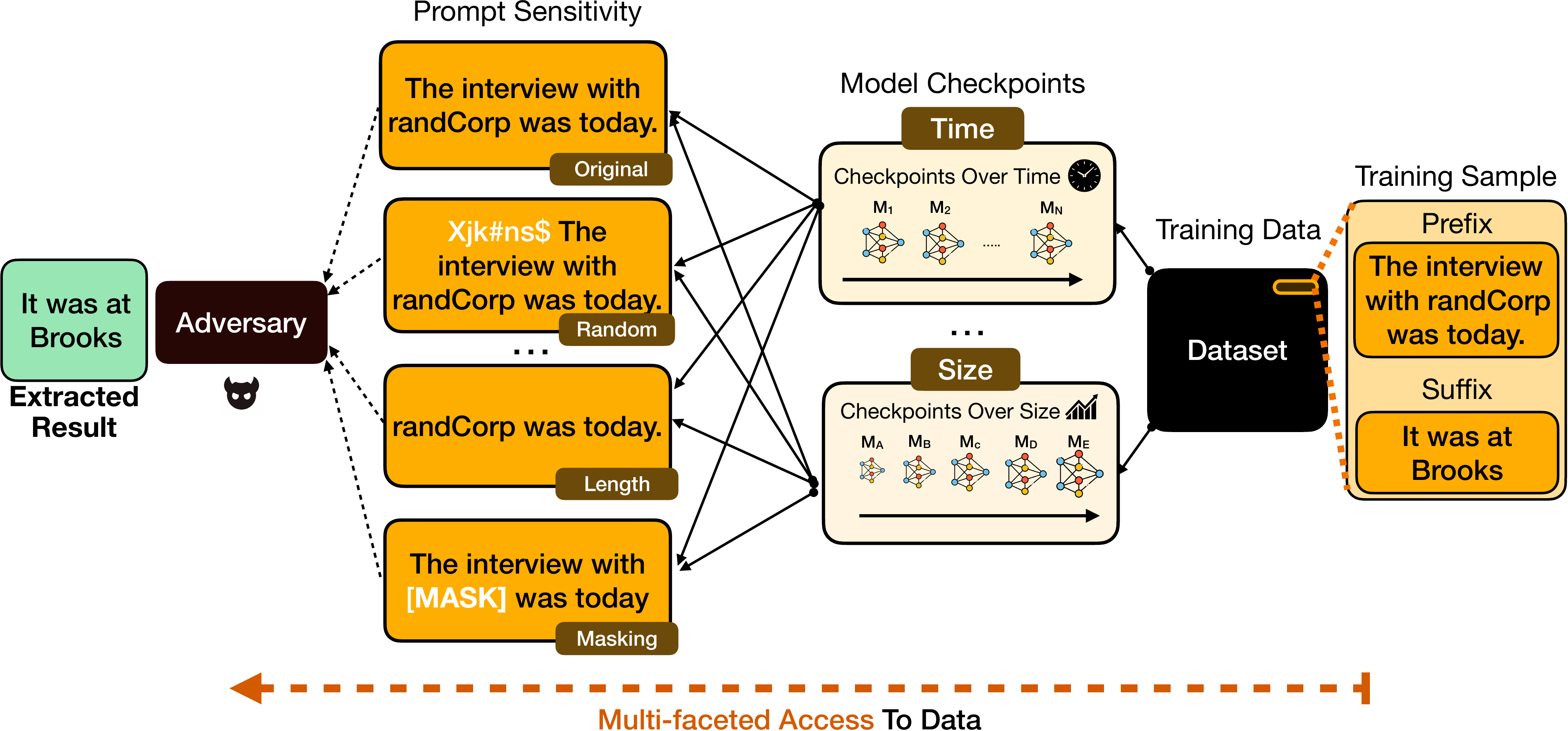}
    \caption{Composability in LLMs. In the real world, an adversary has multi-faceted access to a dataset by (a) exploiting prompt sensitivity, and (b) accessing multiple checkpoints trained on the same data.}
    \label{fig:churn_visualization}
\end{figure*}

\noindent\textbf{Churn.} 
\label{text:churn} 
Churn quantifies the instability of model predictions under updates~\citep{milani2016launch,bahri21a,jiang2021churn,adam2023maintaining,Watson-Daniels2024predictivechurn}, i.e., the inconsistency in predictions between a system pre-update vs post-update, by measuring the fraction of examples whose predictions diverge~\citep{milani2016launch}. While the term churn is traditionally used to describe regressive trends in model predictions, we extend it by highlighting similar regressive trends of extraction attacks under changing prompts and models.
Thus, \textit{churn occurs when information not extractable with a stronger setup is instead extractable with weaker setups like shorter prompts, smaller models, or earlier checkpoints.}
\section{Re-evaluating Adversarial Strengths}
\label{sec:adversarial_strengths}

The adversary is central to our work. We start by defining its capabilities and argue that prior work underestimates real-world adversaries. To ensure broad applicability, we assume gray-box access: the adversary can observe model outputs and probabilities but \textbf{cannot} access weights, gradients, or even control generation hyperparameters, typical in commercial LLMs. Despite these constraints, we show that adversaries are far more powerful than previously recognized due to their multi-faceted access to the underlying data (see Figure \ref{fig:churn_visualization}).

\revisionb{We focus on discoverable memorization, i.e., we assume access to the ground-truth completion to test whether the extracted information is correct. Here, the adversary is primarily interested in auditing the \textit{memorization} behaviour of the model. This is central to many applications, including membership inference, dataset inference, copyright violations, privacy auditing, among others, revisited in \S \ref{sec:case_studies}. That said, as we argue in \S \ref{sec:discussion}, the larger attack surface and implications of a stronger adversary remain even beyond discoverable memorization.}

\subsection{Adversary Capabilities}
\label{sec:adversary_capabilities}

Composability (or self-composability) in privacy~\citep{mcsherry2009privacy} implies that access to multiple outputs from the same data increases the risk of information leakage. Thus, an adversary with multiple access points is much more powerful. In the current landscape of LLMs, such access is not only unsurprising but easily obtainable. Specifically, we focus on two forms of multi-faceted access:

\noindent\textbf{Exploiting Prompt Sensitivity.}
LLMs are highly sensitive to their input, including its structure and content~\citep{scalar.2310.11324,promptingmethods,butterfly.2401.03729, zhu2023promptbench}. While existing studies have focused on improving the prompts for stronger attacks, the nuance of prompt sensitivity in LLMs often defies intuitive expectations. For instance, while longer prompts are known to increase the success of extraction attacks~\cite{carlini2023quantifying}, our work demonstrates that even shorter prompts can at times exploit vulnerabilities that longer prompts overlook (\S \ref{sec:churn_prompt}).

Given the widespread use of LLMs through both chat interfaces and API calls, restricting model access is not realistic. While most commercial LLMs do have rate limits, they are quite high to be of practical concern. For example, even at the lowest tier subscription of $\$5$, ChatGPT has a 500 query per minute (\textit{qpm}) rate limit for GPT4 and 3500 \textit{qpm} for GPT3.5\footnote{\textit{qpm} stats and subscription rate as of September 2024.}. Thus, an adversary can prompt millions of generations in just one day, making it easier to exploit prompt sensitivity.  

\begin{figure*}[t]
    \centering
    \includegraphics[width=0.8\textwidth]{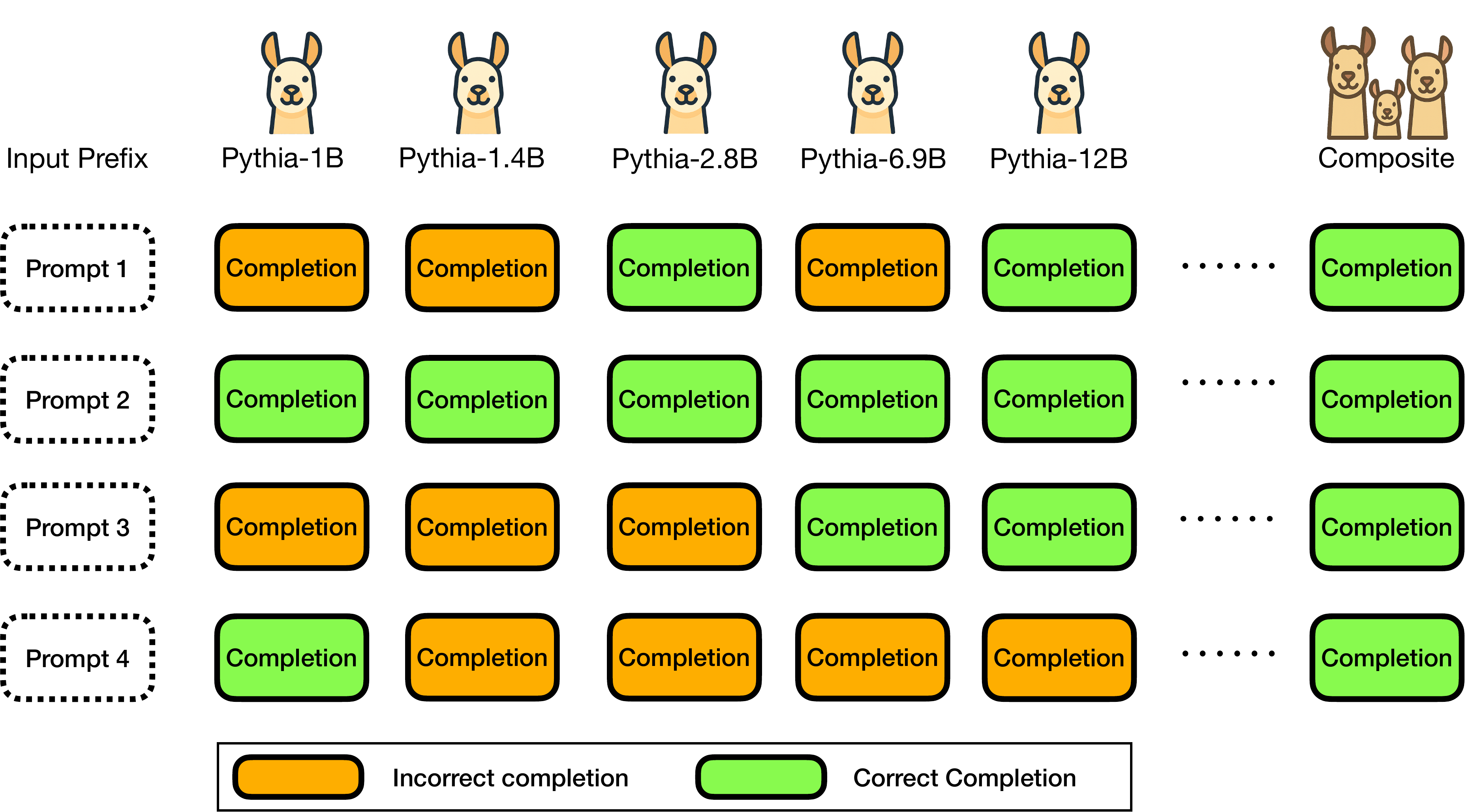}
    \caption{\revision{A toy illustration of how churn may emerge across completions from different model sizes. Adversaries can utilize this churn to increase the number of valid extractions.}}
    \label{fig:churn_minified_example}
\end{figure*}

\noindent\textbf{Multiple Checkpoints.}
LLMs are typically deployed in various sizes to cater to different needs for accuracy and efficiency among users. 
They also undergo regular updates driven by new data, better learning techniques, evolving security measures, and novel functionalities.
Due to the stochastic nature of their training and the impact of scaling, different model sizes or checkpoints might memorize unique portions of the underlying data~\citep{biderman2023emergent}. 
Consequently, an adversary with access to various checkpoints across sizes or training can aggregate extracted information (\S \ref{sec:churn_model}). 


Thus, access to multiple models trained on overlapping datasets substantially increases the attack surface, thereby amplifying the capabilities of adversaries. This level of access has become increasingly common in the current LLM ecosystem. For example, there exist over eight major versions of ChatGPT and ten major versions of Llama, alongside regular minor updates~\cite{chatgpt_modeldocs,chen2023chatgpt}. As such, the availability of multiple models with shared training data can significantly increase the risks of information leakage.

\subsection{Combining Extraction Attacks}
\label{sec:combining}
We discussed the elevated risks of multi-faceted access to the training data. Before presenting our empirical results, we first quantify the risks associated with this stronger real-world adversary. We argue that once the adversary gains such broad access, any successful extraction---even if achieved once---renders that specific information vulnerable to the adversary (see Figure \ref{fig:churn_minified_example}). 

\revisionb{Formally, for a sentence $[\bp\parallel\bx]$ in the training data, where $\parallel$ represents the concatenation of two strings, the adversary has access to the prompt $p$ and aims to extract information $x$.} Adapting the definition of discoverable memorization from \citet{nasr2023scalable}, we propose:
\begin{definition}[\textbf{Composite Discoverable Memorization}]
\label{def:composite_discoverable}
For a set of models $\mathbb{G} = \lbrace {Gen_i|}_{i=1}^k \rbrace$, prompt modifiers $\mathbb{F} = \lbrace {F_j|}_{j=1}^r \rbrace$, and an example $[\bp\parallel\bx]$ from the training dataset $\mathbb{X}$, we say $\bx$ is composite discoverably memorized if $\exists\; Gen_i \in \mathbb{G} \; \textrm{and} \; F_j \in \mathbb{F}$, such that, $Gen_i(F_j(\mathbf{p})) =\mathbf{x}$.
\begin{align*}
    CDM (\mathbb{G}, \mathbb{F}, &\bp \parallel \bx) = \max_{Gen_i \in \mathbb{G}, F_j \in \mathbb{F}} \mathbbm{1}_{Gen_i(F_j(\mathbf{p})) =\mathbf{x}}    
\end{align*}
\end{definition}


Prompt modifiers are defined as functions $F_j: \mathcal{W}^* \rightarrow \mathcal{W}^*$ that take a prompt as input and return a modified version of this prompt as output. Here, $\mathcal{W}$ represents a finite set of all tokens in the training data i.e $\mathcal{W} = \lbrace w_1, w_2, \dots, w_n\rbrace$ with $w_i$ representing individual tokens, and $\mathcal{W}^*$ represents the Kleene star operation over $\mathcal{W}$, i.e., a set of all finite length sequences (strings) of tokens in $\mathcal{W}$.

\begin{figure*}
    \centering
    \includegraphics[width=0.8\textwidth]{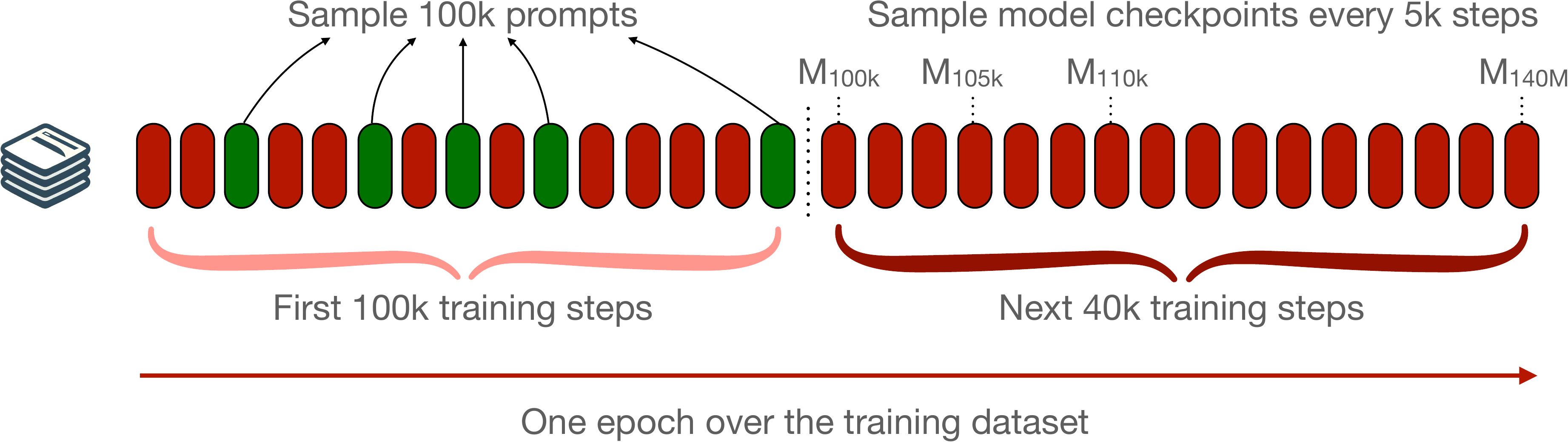}
    \caption{Choosing prompts (pre $100k$ steps) and checkpoints (post $100k$ steps) for evaluation of Pythia.}
    \label{fig:setup_visualization}
\end{figure*}

Extraction attacks are often evaluated in the literature using a verbatim match ~\citep{carlini2021extracting,carlini2023quantifying,nasr2023scalable,huang2022large}, i.e., the generated text must match the original text perfectly. However, this rigid metric does not take into account the noise in LLM generations, and several works have turned to approximate matching~\citep{qi2024follow,kassem2024alpaca,liu2023trustworthy,ippolito2022preventing}. Thus, we also extend our definition of composite extraction attacks to the approximate matching setup as:

\begin{definition}[\textbf{Approximate Composite Discoverable Memorization}]
\label{def:approximate_composite_discoverable}
For a set of models $\mathbb{G} = \lbrace {Gen_i|}_{i=1}^k \rbrace$, prompt modifiers $\mathbb{F} = \lbrace {F_j|}_{j=1}^r \rbrace$, a similarity metric and threshold $S, \delta$, and an example $[\bp\parallel\bx]$ from the training dataset $\mathbb{X}$, we say $\bx$ is approximate composite discoverably memorized if $\exists\; Gen_i \in \mathbb{G} \; \textrm{and} \; F_j \in \mathbb{F}$, such that, $S\left(Gen_i\left(F_j\left(\mathbf{p}\right)\right),\mathbf{x}\right) \geq \delta$.
\begin{align*}
    ACDM (\mathbb{G}, \mathbb{F}, &S, \delta, \bp \parallel \bx) \\ 
    &=  \max_{Gen_i \in \mathbb{G}, F_j \in \mathbb{F}} \mathbbm{1}_{S\left(Gen_i\left(F_j\left(\mathbf{p}\right)\right),\mathbf{x}\right) \geq \delta}
\end{align*}
\end{definition}

Here, $S$ is a similarity metric defined as a function $S: (\mathcal{W}^* \times \mathcal{W}^*) \rightarrow [0, 1]$ that takes as input two strings $a, b \in \mathcal{W}^*$, and returns a score between $0$ and $1$ to represent the similarity between the two input strings, and $\delta$ is a threshold that controls the degree of approximate matching. \revisionb{We experiment with various similarity metrics $S$ in \S \ref{sec:realistic_approximate}}.

\section{Experimental Setup}
\label{sec:setup}

In this section, we outline our central experiment setup to set the stage for our empirical study. 
Further details about the setup for the case studies (\S \ref{sec:case_studies}) are delegated to their respective sections.

\subsection{Models and Dataset}
\label{sec:models_and_datasets}

We use the Pythia suite~\citep{biderman2023pythia} and OLMo models~\citep{groeneveld-etal-2024-olmo} for all our experiments. 
We focus on the Pythia models throughout the paper, while also providing complementary results on OLMo models to show the generalizability of our analysis.
Pythia suite offer access to (a) models of various sizes (1b, 1.4b, 2.8b, 6.9b, and 12b), (b) intermediate checkpoints during training (a total of $154$ checkpoints, with $144$ of them equally spaced, i.e., at every $1k$ training steps), and (c) the training data (Pile dataset~\citep{gao2020pile}) as well as the training order, which is the same for all model sizes. This level of accessibility and control over the training setup allows us to simulate the real-world availability of models across sizes and different checkpoints over time. 

OLMo models were trained on the Dolma dataset~\cite{soldaini-etal-2024-dolma} and also offer access to (a) intermediate model checkpoints during training, and (b) the complete training data order.

\subsection{Evaluation Methodology}

We now describe our evaluation methodology. Similar to \citet{carlini2023quantifying}, we sample a representative portion of the training data for analyzing the performance of our extraction attacks. More specifically, we uniformly sample $100,000$ sequences (prompts) from the first $100k$ steps (batches) of the training data for Pythia. This sampling strategy is important because we choose model checkpoints for evaluation starting at step $100k$, which ensures that every sentence evaluated for memorization has been seen by each checkpoint under consideration (as illustrated in Figure \ref{fig:setup_visualization}).

Each sequence sampled is exactly $2049$ tokens. For our analysis, we employ a consistent method of partitioning each sequence into a prompt and completion at the midpoint, i.e., $1024$ tokens. Formally, for a sentence $s_{1:2049}$, prompt length $l_p$, and completion length $l_x$, the example $[\bp\parallel\bx]$ is defined as $\bp = s_{1024-l_p:1024}$ and $\bx = s_{1024:1024+l_x}$. 
This partitioning allows us to systematically vary the prompt length and design while comparing the same completion, and vice versa. 
We use the same approach for OLMo, with the training step $300k$ (instead of $100k$) being the cut-off point.

For the Pythia suite, unless otherwise specified, we use a prompt length of $l_p=50$, a completion length of $l_x=50$, the Pythia-6.9b model, and the 140k training step checkpoint, evaluating the extraction attacks using verbatim match.
For OLMo, we use the OLMo-7b model and the 500k training step checkpoint as defaults, while the rest of the configuration is the same as Pythia.




\section{Churn in Extraction Trends}
\label{sec:churn}

\begin{figure*}[t!]
    \centering
    \begin{tikzpicture}
\begin{axis} [width=0.32\linewidth,height=0.27\linewidth,yticklabel= {\pgfmathprintnumber\tick\%},ymax=4.2,ymin=0.,ylabel={Extraction Rate},xlabel={Prompt Length},enlarge x limits=0.05,axis y line*=left,axis x line*=bottom,ylabel near ticks,ylabel style={align=center, font=\small},xlabel style={align=center, text width=4cm, font=\small},
xtick={50, 200, 350, 500},ticklabel style = {font=\small}
,title style={align=center, text width=4cm, font=\small},
title={\textbf{(a) Across Prompt Length}}
]

\path[name path=base] (axis cs:50,0) -- (axis cs:500,0);

\addplot [name path=composite,darkgreenvar,ultra thick,dashed] coordinates {(50,3.514) (500,3.514)};
\node[anchor=south west,font=\small,darkgreenvar] at (axis cs: 50,3.514) {Composite Ext. Rate};

\addplot [name path=all,chromevar,ultra thick,dotted,mark=*,mark options={scale=1,solid}] table[x=promptlen,y=acc] {data/promptlen_exact_overlap_complen50.dat};

\addplot [color=chromevar,fill=chromevar, fill opacity=0.3] fill between[of=all and base];
\addplot [color=darkgreenvar,fill=darkgreenvar, fill opacity=0.1] fill between[of=all and composite];

\end{axis}
\end{tikzpicture}%
\hfill
\begin{tikzpicture}
\begin{axis} [width=0.32\linewidth,height=0.27\linewidth,yticklabel= {\pgfmathprintnumber\tick\%},ymax=2,ymin=0.,xlabel={Model Size \vphantom{p}},enlarge x limits=0.05,axis y line*=left,axis x line*=bottom,ylabel near ticks,ylabel style={align=center, text width=4cm, font=\small},xlabel style={align=center, text width=4cm, font=\small},symbolic x coords={{1b},{1.4b},{2.8b},{6.9b},{12b}},
xtick={1b, 1.4b, 2.8b, 6.9b, 12b},ticklabel style = {font=\small},
,title style={align=center, text width=4cm, font=\small},
title={\textbf{(b) Across Model Size \vphantom{p}}}
]

\path[name path=base] (axis cs:1b,0) -- (axis cs:12b,0);

\addplot [name path=composite,darkgreenvar,ultra thick,dashed] coordinates {(1b,1.614) (12b,1.614)};
\node[anchor=south west,font=\small,darkgreenvar] at (axis cs: 1b,1.614) {Composite Ext. Rate};

\addplot [name path=all,chromevar,ultra thick,dotted,mark=*,mark options={scale=1,solid}] table[x=modelsize,y=acc] {data/modelsize_exact_overlap_complen50.dat};

\addplot [color=chromevar,fill=chromevar, fill opacity=0.3] fill between[of=all and base];
\addplot [color=darkgreenvar,fill=darkgreenvar, fill opacity=0.1] fill between[of=all and composite];

\end{axis}
\end{tikzpicture}%
\hfill
\begin{tikzpicture}
\begin{axis} [width=0.32\linewidth,height=0.27\linewidth,yticklabel= {\pgfmathprintnumber\tick\%},ymax=2.,ymin=0.,xlabel={Training Step},enlarge x limits=0.05,axis y line*=left,axis x line*=bottom,ylabel near ticks,ylabel style={align=center, text width=4cm, font=\small},xlabel style={align=center, text width=4cm, font=\small},symbolic x coords={{100k},{105k},{110k},{115k},{120k},{125k},{130k},{135k},{140k}},
xtick={100k, 110k, 120k, 130k, 140k},ticklabel style = {font=\small},
,title style={align=center, text width=4cm, font=\small},
title={\textbf{(c) Across Training Steps}}
]

\path[name path=base] (axis cs:100k,0) -- (axis cs:140k,0);

\addplot [name path=composite,darkgreenvar,ultra thick,dashed] coordinates {(100k,1.638) (140k,1.638)};
\node[anchor=south west,font=\small,darkgreenvar] at (axis cs: 100k,1.638) {Composite Ext. Rate};

\addplot [name path=all,chromevar,ultra thick,dotted,mark=*,mark options={scale=1,solid}] table[x=modelstep,y=acc] {data/modelstep_exact_overlap_complen50.dat};

\addplot [color=chromevar,fill=chromevar, fill opacity=0.3] fill between[of=all and base];
\addplot [color=darkgreenvar,fill=darkgreenvar, fill opacity=0.1] fill between[of=all and composite];

\end{axis}
\end{tikzpicture}

\begin{tikzpicture}
\begin{axis}[width=0.33\linewidth,height=0.24\linewidth,yticklabel= {\pgfmathprintnumber\tick\%},ymin=0,ylabel={Extraction Rate},xlabel={Prompt Length}, enlarge x limits=0.25,xtick=data,ybar=0.5pt,bar width=4pt,axis y line*=left,axis x line*=bottom,symbolic x coords={{100},{300},{500}},ylabel near ticks,ylabel style={align=center, text width=4cm, font=\small},xlabel style={align=center, text width=4cm, font=\small},
xtick={100, 300, 500},ticklabel style = {font=\small},
,title style={align=center, text width=4cm, font=\small},
title={\textbf{(d) Masking and Removal}}
]

\addplot[style={fill=sapgreenvar},draw=none
]
coordinates {({100},1.831)
             ({300},2.835)
	 	({500},3.2009999999999996)
	     };
\addplot[style={fill=lightbrownvar},draw=none
]
coordinates {({100},0.506)
             ({300},0.627)
	 	({500},0.632)
	     };
\addplot[style={fill=darkbrownvar},draw=none
]
coordinates {({100},0.186)
             ({300},0.282)
	 	({500},0.146000000000000004)
	     };
\addplot[style={fill=lightvioletvar},draw=none
]
coordinates {({100},0.86)
             ({300},1.052)
	 	({500},1.043)
	     };
\addplot[style={fill=darkvioletvar},draw=none
]
coordinates {({100},0.387)
             ({300},0.51299999999999999)
	 	({500},0.51399999999999999)
	     };
\addplot[style={fill=darkgreenvar},draw=none
]
coordinates {({100},1.854)
             ({300},2.906)
	 	({500},3.273)
	     };

\end{axis}
\end{tikzpicture}%
\raisebox{3em}{
\begin{tikzpicture}
\begin{axis}[scale=0.01,
legend cell align={left},
hide axis,
xmin=0, xmax=1,
ymin=0, ymax=1,
legend columns=1,
legend style={font=\footnotesize,/tikz/every even column/.append style={column sep=0.1cm}},
legend image code/.code={
        \draw [#1] (0cm,-0.05cm) rectangle (0.3cm,0.1cm); },
]
]

\addlegendimage{ultra thick, sapgreenvar, fill=sapgreenvar}
\addlegendentry{Baseline};
\addlegendimage{ultra thick, lightbrownvar, fill=lightbrownvar}
\addlegendentry{Mask ($5\%$)};
\addlegendimage{ultra thick, darkbrownvar, fill=darkbrownvar}
\addlegendentry{Mask ($10\%$)};
\addlegendimage{ultra thick, lightvioletvar, fill=lightvioletvar}
\addlegendentry{Remove ($5\%$)};
\addlegendimage{ultra thick, darkvioletvar, fill=darkvioletvar}
\addlegendentry{Remove ($10\%$)};
\addlegendimage{ultra thick, darkgreenvar, fill=darkgreenvar}
\addlegendentry{Composite};

\end{axis}
\end{tikzpicture}
}%
\hfill
\begin{tikzpicture}
\begin{axis}[width=0.2\linewidth,height=0.24\linewidth,yticklabel= {\pgfmathprintnumber\tick\%},ymin=0,ylabel={Extraction Rate},xlabel={{\color{white}Prefix \vphantom{g}}},enlarge x limits=0.25,xtick=data,ybar=0.5pt,bar width=4pt,axis y line*=left,axis x line*=bottom,symbolic x coords={{Prefix}},ylabel near ticks,ylabel style={align=center, text width=4cm, font=\small},xlabel style={align=center,font=\small},ticklabel style = {font=\small},
,title style={align=center, text width=3.5cm, font=\small},
title={\textbf{(e) Random Prefixes \vphantom{g}}}
]
s

\addplot[style={fill=sapgreenvar},draw=none
]
coordinates {({Prefix},1.258)
	     };
\addplot[style={fill=lightbluevar},draw=none
]
coordinates {({Prefix},0.831)
	     };
\addplot[style={fill=crazybluevar},draw=none
]
coordinates {({Prefix},0.631)
	     };
\addplot[style={fill=darkgreenvar},draw=none
]
coordinates {({Prefix},1.295)
	     };

\end{axis}
\end{tikzpicture}%
\raisebox{3em}{
\begin{tikzpicture}[trim left=-4.4em]
\begin{axis}[scale=0.01,
legend cell align={left},
hide axis,
xmin=0, xmax=1,
ymin=0, ymax=1,
legend columns=1,
legend style={font=\footnotesize,/tikz/every even column/.append style={column sep=0.1cm}},
legend image code/.code={
        \draw [#1] (0cm,-0.05cm) rectangle (0.3cm,0.1cm); },
]
]

\addlegendimage{ultra thick, sapgreenvar, fill=sapgreenvar}
\addlegendentry{No Prefix};
\addlegendimage{ultra thick, lightbluevar, fill=lightbluevar}
\addlegendentry{Numbers};
\addlegendimage{crazybluevar,fill=crazybluevar,postaction={pattern=horizontal lines}}
\addlegendentry{Alphanumeric};
\addlegendimage{ultra thick, darkgreenvar, fill=darkgreenvar}
\addlegendentry{Composite};
\end{axis}
\end{tikzpicture}
}
    \caption{Extraction rates under prompt sensitivity and across models for Pythia. \textbf{(a)} Increasing prompt length improves extraction, with the composite extraction rate better than even at prompt length $500$. 
    \textbf{(b, c)} Increasing model size and training steps show similar trends, with the largest impact of the composite extraction rate seen in training steps, increasing the extraction rate $1.5 \times$ compared to a single checkpoint.
    \textbf{(d)} Randomly masking or removing tokens severely hurts the extraction rate, highlighting the importance of prompt structure. 
    \textbf{(e)} Adding a random prefix can also contribute to the composite extraction rate.}
    \label{fig:churn}
\end{figure*}

Churn~\citep{milani2016launch}, as previously introduced in \S \ref{sec:related_work}, refers to regressive variance for individual extracted information despite an overall improvement in the extraction rates. 
For instance, although using a longer prompt is often associated with stronger extraction rates~\citep{carlini2023quantifying,biderman2023emergent}, we observe trends that exhibit churn, i.e., certain information is instead extractable only with shorter prompts but not with longer prompts.
These non-monotonic and locally regressive trends of certain sentences (i.e., churn) can be exploited by an adversary with multifaceted access to the data to execute a composite extraction attack.
We study the factors that may lead to \textit{churn} such as (a) prompt sensitivity, and (b) access to models of varying sizes and training checkpoints.

\subsection{Prompt Sensitivity}
\label{sec:churn_prompt}

We start by examining prompt sensitivity, focusing on how trends in prompt design can lead to churn.

\noindent\textbf{Prompt Length.}
Prompt length is a commonly studied parameter in extraction attacks, and it has been shown that longer prompts lead to better extraction~\citep{carlini2023quantifying}.
This is intuitive, as conditioning the model with more text from training would increase the likelihood of extraction. 

However, we show in Figure \ref{fig:churn}(a) that the composite extraction rate (Definition \ref{def:composite_discoverable}) across varying prompt lengths is noticeably higher than the extraction rate at even the largest prompt length at 500 tokens. 
Thus, certain information extractable with shorter prompts remains elusive even with the longest prompt, \revisionb{due to prompt sensitivity in LLMs, as discussed in \S \ref{sec:adversary_capabilities}.}
Consequently, an adversary can exploit this churn across the prompt length to extract more information. We see similar trends for OLMo in Figure \ref{fig:olmo}.


\begin{figure}[t!]
    \centering
    \vspace{-1.7em}
    \begin{tikzpicture}
\begin{axis}[width=0.52\linewidth,height=0.5\linewidth,yticklabel= {\pgfmathprintnumber\tick\%},ymin=0,xlabel={},ylabel={Extraction Rate}, enlarge x limits=0.45,xtick=data,ybar=0.5pt,bar width=4pt,axis y line*=left,axis x line*=bottom,symbolic x coords={{Prompt Length},{Training Steps}},ylabel near ticks,ylabel style={align=center, text width=4cm, font=\small},xlabel style={align=center, text width=4cm, font=\small},
xtick={Prompt Length, Training Steps},ticklabel style = {font=\small},xticklabel style = {text width=0.9cm}
]

\addplot[style={fill=sapgreenvar},draw=none
]
coordinates {({Prompt Length},0.5108476286579213)
             ({Training Steps},0.4372687521022536)
	     };
\addplot[style={fill=lightbluevar},draw=none
]
coordinates {({Prompt Length},2.064413050790447)
             ({Training Steps},0.5108476286579213)
	     };
\addplot[style={fill=darkgreenvar},draw=none
]
coordinates {({Prompt Length},2.37989404641776)
             ({Training Steps},0.8251345442314161)
	     };

\end{axis}
\end{tikzpicture}%
\hspace{1em}
\raisebox{2em}{
\begin{tikzpicture}
\begin{axis}[scale=0.01,
legend cell align={left},
hide axis,
xmin=0, xmax=1,
ymin=0, ymax=1,
legend columns=1,
legend style={font=\footnotesize,/tikz/every even column/.append style={column sep=0.1cm}},
legend image code/.code={
        \draw [#1] (0cm,-0.05cm) rectangle (0.3cm,0.1cm); },
]
]

\addlegendimage{ultra thick, sapgreenvar, fill=sapgreenvar}
\addlegendentry{Prompt Length $50$};
\addlegendimage{ultra thick, lightbluevar, fill=lightbluevar}
\addlegendentry{Prompt Length $500$};
\addlegendimage{ultra thick, white, fill=white}
\addlegendentry{};
\addlegendimage{ultra thick, sapgreenvar, fill=sapgreenvar}
\addlegendentry{Training Step $300k$};
\addlegendimage{ultra thick, lightbluevar, fill=lightbluevar}
\addlegendentry{Training Step $500k$};
\addlegendimage{ultra thick, white, fill=white}
\addlegendentry{};
\addlegendimage{ultra thick, darkgreenvar, fill=darkgreenvar}
\addlegendentry{Composite All};

\end{axis}
\end{tikzpicture}
}%
    \caption{Composite extraction attack results across 10 prompt lengths (same as Pythia) and 11 training steps (equidistant between $300k$ and $500k$), compared against isolated setups, for OLMo.
    }
    \label{fig:olmo}
\end{figure}

\begin{figure*}[h]
    \centering
    \includegraphics[width=0.48\textwidth]{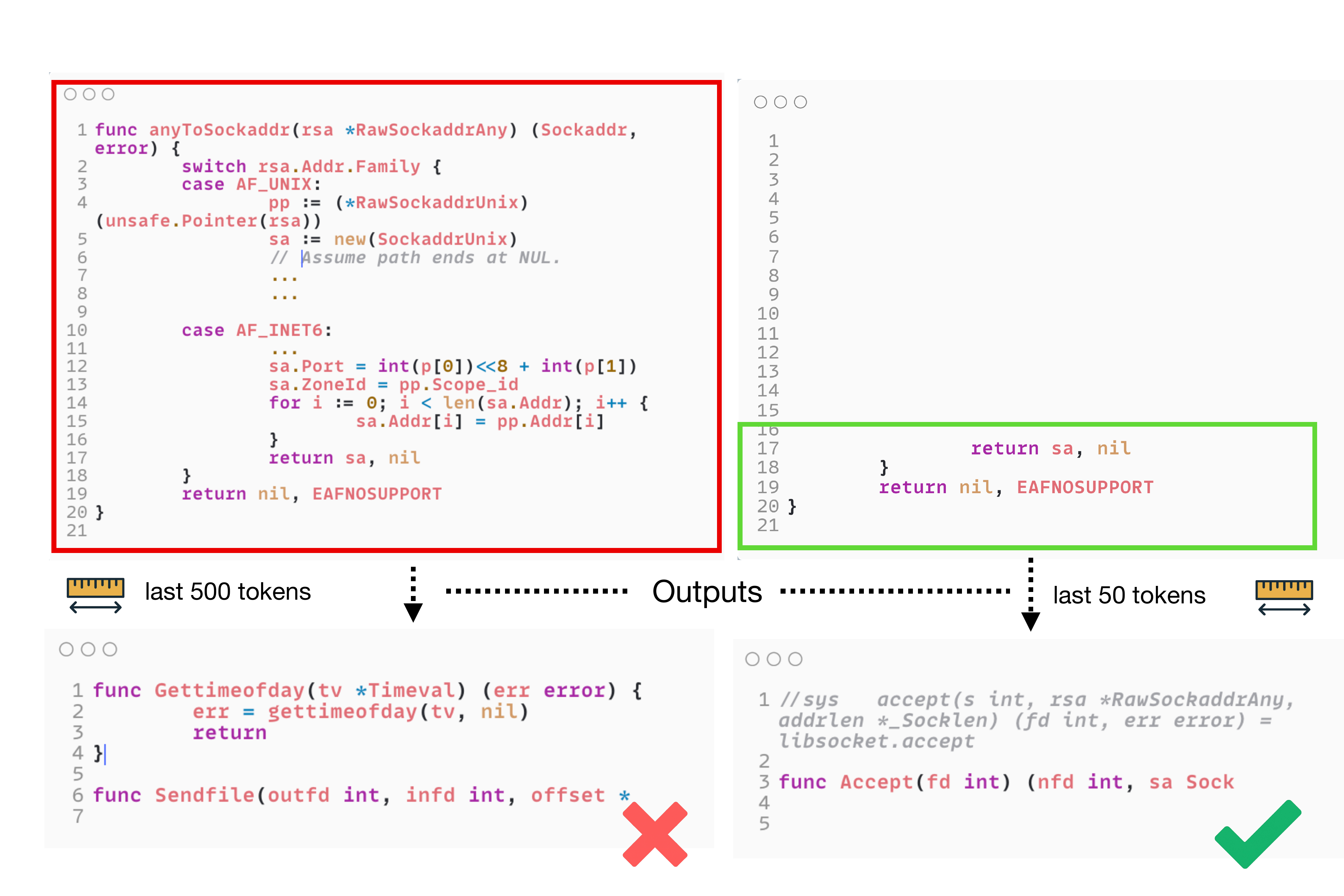}%
    \hfill
    \includegraphics[width=0.5\textwidth]{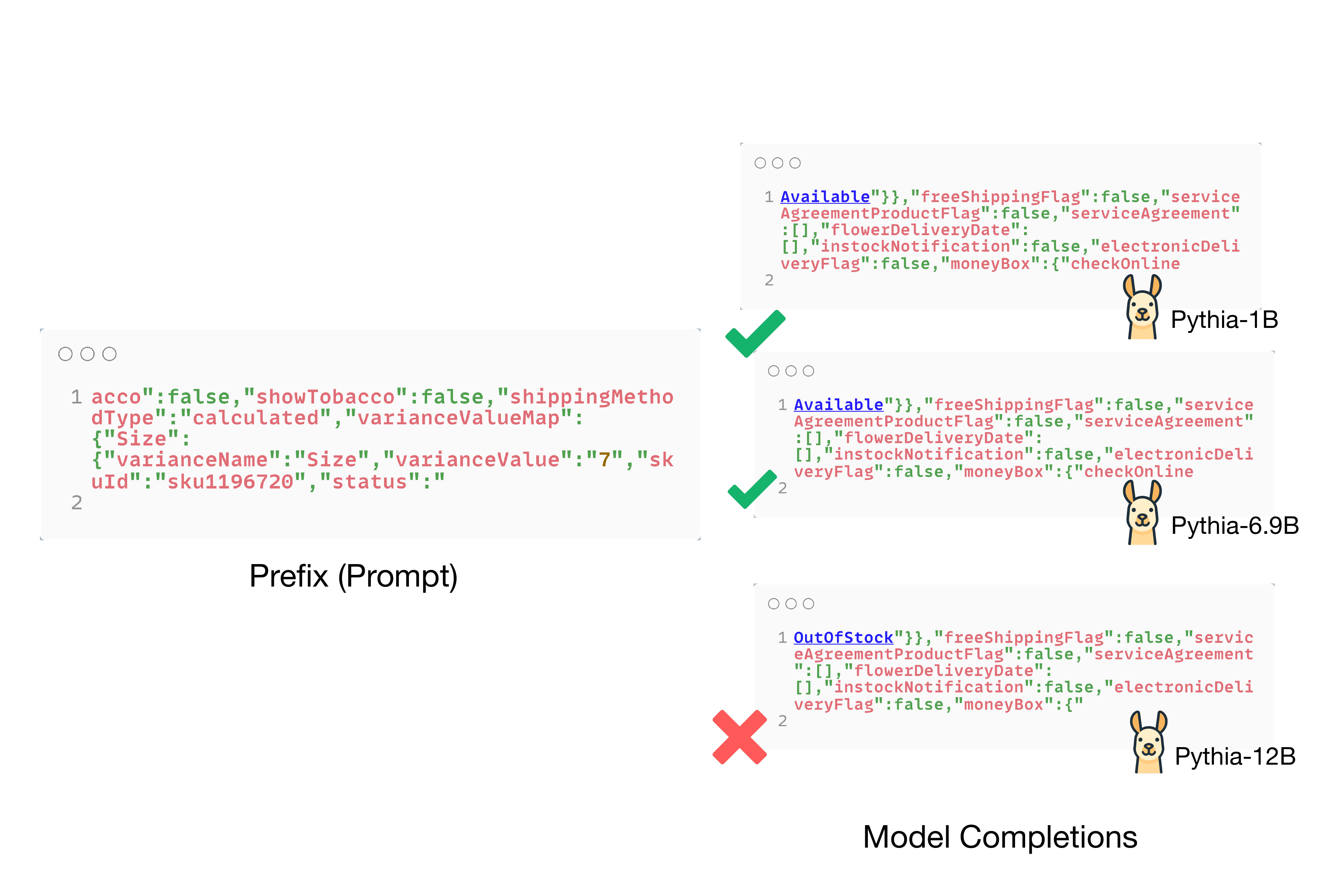}
    \caption{\revisionb{Real examples of churn. \textbf{(Left)} Prompts of different lengths can contribute uniquely to the model extraction, and longer contexts aren't always better. \textbf{(Right)} Similarly, different model sizes contribute uniquely to the composite extraction rate, showing the importance of churn across model size.}}
    \label{fig:churn_smaller_length_produces_more}
\end{figure*}

\noindent\textbf{Prompt Structure.} Next, we explore the structure of prompts to identify where churn can emerge. We introduce noise into the prompts by masking and removing random tokens (Figure \ref{fig:churn}(d)); and as a prefix in the form of random numeric and alphanumeric strings (Figure \ref{fig:churn}(e)). Despite introducing only a small amount of noise, we observe a significant drop in extraction rates. This indicates that the contiguous prompt from the training data is crucial for extracting information, and any disruption to this prompt can significantly hurt its capabilities. 

Yet, we do see churn in extraction trends, with a larger impact of the noisy prefixes. This further highlights how an adversary can exploit repeated prompting, even with seemingly unintuitive changes like masking or removing random tokens and adding a random prefix to the input prompt.

Note that the churn trends under prompt sensitivity, \revisionb{both for prompt structure and prompt length,} highlight the increased extraction risks without access to new information. For instance, if an adversary has access to the prompt of length 500 tokens, they can expand the attack surface and thereby the extraction rate simply by removing parts of the prompt \revisionb{(prompt length composite attack from Figure \ref{fig:churn}(a))}, adding noise \revisionb{(prompt structure composite attack from Figure \ref{fig:churn}(d, e))}, etc., without needing any additional knowledge. 

One might argue that as the number of prompt variations increases, every sentence could become extractable. However, that is not true; not all sentences are extractable. \citet{yin2024benchmarking} showed that knowledge not present in an LLM will not be extractable even after prompt optimization, while \citet{schwarzschild2024rethinking} also showed similar trends when attempting to extract a given completion. 
Consequently, prompting an LLM to regurgitate certain sentences, even with various prompt modifications, demonstrates a genuine extraction risk and underscores the extent of memorization in LLMs \citep{carlini2021extracting}.

\begin{figure*}[t!]
    \centering
    \begin{tikzpicture}
\begin{axis}[width=0.32\linewidth,height=0.24\linewidth,yticklabel= {\pgfmathprintnumber\tick\%},ymin=0,ylabel={Extraction Rate},xlabel={Number of Models}, enlarge x limits=0.25,xtick=data,ybar=0.5pt,bar width=10pt,axis y line*=left,axis x line*=bottom,symbolic x coords={1, 2, 3, 4, 5},ylabel near ticks,ylabel style={align=center, text width=4cm, font=\small},xlabel style={align=center, text width=4cm, font=\small},
xtick={1, 2, 3, 4, 5},ticklabel style = {font=\small},
,title style={align=center, text width=4cm, font=\small},
title={\textbf{(a) Recurrence of Extractions across Model Size}}
]

\addplot[style={fill=darkgreenvar},draw=none
]
coordinates {(1,0.357)
             (2,0.232)
	 	(3,0.204)
            (4,0.193)
            (5,0.628)
	     };

\end{axis}
\end{tikzpicture}
\begin{tikzpicture}
\begin{axis}[width=0.32\linewidth,height=0.24\linewidth,yticklabel= {\pgfmathprintnumber\tick\%},ymin=0,xlabel={Model Size}, enlarge x limits=0.15,xtick=data,ybar=0.5pt,bar width=10pt,axis y line*=left,axis x line*=bottom,symbolic x coords={{1b},{1.4b},{2.8b},{6.9b},{12b}},ylabel near ticks,ylabel style={align=center, text width=4cm, font=\small},xlabel style={align=center, text width=4cm, font=\small},
xtick={{1b},{1.4b},{2.8b},{6.9b},{12b}},ticklabel style = {font=\small},
,title style={align=center, text width=4cm, font=\small},
title={\textbf{(b) Unique Extractions across Model Size}}
]

\addplot[style={fill=darkgreenvar},draw=none
]
coordinates {({1b},0.013)
             ({1.4b},0.014)
	 	({2.8b},0.026)
            ({6.9b},0.093)
            ({12b},0.211)
	     };

\end{axis}
\end{tikzpicture}
\begin{tikzpicture}
\begin{axis}[width=0.32\linewidth,height=0.24\linewidth,yticklabel= {\pgfmathprintnumber\tick\%},ymin=0,xlabel={Model Size}, enlarge x limits=0.15,xtick=data,ybar=0.5pt,bar width=10pt,axis y line*=left,axis x line*=bottom,symbolic x coords={{1b},{1.4b},{2.8b},{6.9b},{12b}},ylabel near ticks,ylabel style={align=center, text width=4cm, font=\small},xlabel style={align=center, text width=4cm, font=\small},
xtick={{1b},{1.4b},{2.8b},{6.9b},{12b}},ticklabel style = {font=\small},
,title style={align=center, text width=4cm, font=\small},
title={\textbf{(c) Drop-One Extractions across Model Size}}
]

\addplot[style={fill=darkgreenvar},draw=none
]
coordinates {({1b},1.601)
             ({1.4b},1.6)
	 	({2.8b},1.588)
            ({6.9b},1.521)
            ({12b},1.403)
	     };

\end{axis}
\end{tikzpicture}
    \caption{\revision{Granular trends of churn across model size for Pythia. \textbf{(a)} Number of models that successfully extract the same information, i.e., whether only 1 out of 5 models extracts a given sample, or whether 2, 3, 4, or all 5 do. (The x-axis here indicates the exact number of models that extract the same samples.)
    Even though the majority of samples are extractable by all 5 models, a significant amount of extractions are unique to a subset, or even a single model. \textbf{(b)} While larger models contribute more unique extractions, each model, regardless of size, provides some unique extractions.
    \textbf{(c)} Composite extraction rates after dropping a single model (x-axis: dropped model, y-axis: resulting extraction rate) show that extraction rates remain stable even if we remove the contributions of the biggest model.}}
    \label{fig:unique}
\end{figure*}

\subsection{Multiple Checkpoints}
\label{sec:churn_model}

\noindent\textbf{Model Size.}
The model size has long been known to influence learning trends, and our results in Figure \ref{fig:churn}(b) reflect this phenomenon. Larger models tend to memorize more information, which makes them more vulnerable to extraction attacks. However, our results also indicate that the composite extraction rate is higher than the extraction rate of any single model, highlighting the churn in these trends. \citet{biderman2023emergent} also conducted an empirical study on the overlap between memorized data across model sizes and found that up to $10\%$ data memorized by smaller models is not memorized by larger models. Combining our insights with existing literature, it's clear that releasing models in different sizes increases the extraction risks.

\noindent\textbf{Model Updates.}
We also analyze model updates over time using intermediate checkpoints in Figure \ref{fig:churn}(c), where we observe the most significant churn in our study. Unsurprisingly, attacking models at later stages of training is more successful, as seen in the literature~\citep{tirumala2022memorization,biderman2023emergent,jagielskimeasuring}. But remarkably, the churn here is significantly powerful and by exploiting composability, an adversary can increase their extraction rate by more than $1.5 \times$. This underscores the impact of stochasticity in model training on extraction attacks and reveals that regular model updates, typically considered beneficial in the LLM ecosystem, create a powerful adversary.
We also see similar results for OLMo in Figure \ref{fig:olmo}.

\revision{We provide some examples of extractions from the Pile dataset that show regressive trends, i.e., successful extraction using weaker setups, and highlight the value of churn, in Figure \ref{fig:churn_smaller_length_produces_more}.}


\subsection{Unique Extractions}

\revision{Next, we study the contribution of each setup to the composite extraction rate. We focus on trends across model size, while providing results for other axes in the appendix (\S \ref{sec:app_unique}). We find that while a majority of extracted samples are extractable with all 5 model sizes, there is a significant portion of extractions unique to a few (or even just one) model(s) (Figure \ref{fig:unique}(a)). Examining the distribution of extractions unique to individual models (Figure \ref{fig:unique}(b)), we observe that Pythia-6.9b produces a substantial number of unique extractions not found even in the larger Pythia-12b. Smaller models, such as Pythia-1b, 1.4b, and 2.8b, also contribute non-trivially.}

\revision{Finally, we study composite extraction rates when all models except one are used (Figure \ref{fig:unique}(c)), to quantify the contribution of each model. 
We find that even after dropping the largest model, the remaining models achieve high composite extraction rates, indicating that no single model is essential for strong extraction. Given that attacking the most vulnerable model can be expensive, this shows that an adversary can take advantage of the churn to instead perform a composite attack on less vulnerable but cheaper models (more details in \S \ref{sec:realistic_combine}).} 

\section{Towards Realistic Extraction Attacks}
\label{sec:realistic}

With a better understanding of the churn, we now evaluate a more realistic measure of leakage in extraction attacks, by investigating (a) composability across multiple axes, (b) cost of composite attacks, (c) approximate matching, and (d) deduplication.

\begin{figure*}[t!]
    \centering
    \begin{tikzpicture}[transform shape]
    \node[font=\small] at (0.,0.) {\textbf{(a) Composite Extraction Rates Across Multiple Axes}};
\end{tikzpicture}

\begin{tikzpicture}[inner sep=0., scale=0.8, transform shape]
    \foreach \y [count=\n] in {{1.19, 1.53, 1.85}, {1.71, 2.11, 2.50}, {2.03, 2.49, 2.90}}{
        \foreach \x [count=\m] in \y {
            \pgfmathsetmacro \clr {\x*50}
            \pgfmathsetmacro \ns {\n*1.0}
            \pgfmathsetmacro \ms {\m*1.2}
            \node[fill=chromevar!\clr, minimum width=12mm, minimum height=10mm] at (\ms,-\ns) {\small\x\%};
        }
    }

    \node[rotate=90] at (-0.6, -2.1) {Prompt Length};
    \node[] at (2.4, 0.5) {Model Size};
    \foreach \xlabel [count=\m] in {1.4b,+2.8b,+6.9b}{
        \pgfmathsetmacro \ms {\m*1.2};
        \node[minimum width=12mm, minimum height=10mm] at (\ms,0) {\xlabel};
    }
    \foreach \xlabel [count=\m] in {100,+200,+300}{
        \pgfmathsetmacro \ms {\m*1.2};
        \node[minimum width=12mm, minimum height=10mm,rotate=90] at (0,-\m) {\xlabel};
    }
\end{tikzpicture}%
\hspace{1.5em}
\begin{tikzpicture}[inner sep=0., scale=0.8, transform shape]
    \foreach \y [count=\n] in {{1.19, 1.71, 2.03}, {1.41, 1.96, 2.29}, {1.52, 2.08, 2.42}}{
        \foreach \x [count=\m] in \y {
            \pgfmathsetmacro \clr {\x*50}
            \pgfmathsetmacro \ns {\n*1.0}
            \pgfmathsetmacro \ms {\m*1.2}
            \node[fill=chromevar!\clr, minimum width=12mm, minimum height=10mm] at (\ms,-\ns) {\small\x\%};
        }
    }

    \node[rotate=90] at (-0.6, -2.1) {Training Steps};
    \node[] at (2.4, 0.5) {Prompt Length};
    \foreach \xlabel [count=\m] in {100,+200,+300}{
        \pgfmathsetmacro \ms {\m*1.2};
        \node[minimum width=12mm, minimum height=10mm] at (\ms,0) {\xlabel};
    }
    \foreach \xlabel [count=\m] in {100k,+120k,+140k}{
        \pgfmathsetmacro \ms {\m*1.2};
        \node[minimum width=12mm, minimum height=10mm,rotate=90] at (0,-\m) {\xlabel};
    }
\end{tikzpicture}%
\hspace{1.5em}
\begin{tikzpicture}[inner sep=0., scale=0.8, transform shape]
    \foreach \y [count=\n] in {{1.19, 1.41, 1.52}, {1.53, 1.76, 1.87}, {1.85, 2.12, 2.27}}{
        \foreach \x [count=\m] in \y {
            \pgfmathsetmacro \clr {\x*50}
            \pgfmathsetmacro \ns {\n*1.0}
            \pgfmathsetmacro \ms {\m*1.2}
            \node[fill=chromevar!\clr, minimum width=12mm, minimum height=10mm] at (\ms,-\ns) {\small\x\%};
        }
    }

    \node[rotate=90] at (-0.6, -2.1) {Model Size};
    \node[] at (2.4, 0.5) {Training Steps};
    \foreach \xlabel [count=\m] in {100k,+120k,+140k}{
        \pgfmathsetmacro \ms {\m*1.2};
        \node[minimum width=12mm, minimum height=10mm] at (\ms,0) {\xlabel};
    }
    \foreach \xlabel [count=\m] in {1.4b,+2.8b,+6.9b}{
        \pgfmathsetmacro \ms {\m*1.2};
        \node[minimum width=12mm, minimum height=10mm,rotate=90] at (0,-\m) {\xlabel};
    }
\end{tikzpicture}

\begin{tikzpicture}
\begin{axis}[width=0.3\linewidth,height=0.26\linewidth,yticklabel= {\pgfmathprintnumber\tick\%},ymin=0, ylabel={Extraction Rate},xlabel={Dataset}, enlarge x limits=0.6,xtick=data,ybar=0.5pt,bar width=4pt,axis y line*=left,axis x line*=bottom,symbolic x coords={{Pile},{Pile-Deduped}},ylabel near ticks,ylabel style={align=center, text width=4cm, font=\small},xlabel style={align=center, text width=4cm, font=\small},
xtick={Pile, Pile-Deduped},ticklabel style = {font=\small},
,title style={align=center, text width=6.5cm, font=\small},
title={\textbf{(b) \vphantom{g}Composite Attacks and Deduplication\vphantom{g}}}
]

\addplot[style={fill=sapgreenvar},draw=none
]
coordinates {({Pile},1.1860000000000002)
             ({Pile-Deduped},0.941)
	     };

\addplot[style={fill=lightbluevar},draw=none
]
coordinates {({Pile},2.838)
             ({Pile-Deduped},2.1950000000000003)
	     };
\addplot[style={fill=darkgreenvar},draw=none
]
coordinates {({Pile},3.447)
             ({Pile-Deduped},2.732)
	     };

\end{axis}
\end{tikzpicture}%
\raisebox{2.5em}{
\begin{tikzpicture}[trim left=-2em]
\begin{axis}[scale=0.01,
legend cell align={left},
hide axis,
xmin=0, xmax=1,
ymin=0, ymax=1,
legend columns=1,
legend style={font=\scriptsize,/tikz/every even column/.append style={column sep=0.1cm}},
legend image code/.code={
        \draw [#1] (0cm,-0.05cm) rectangle (0.3cm,0.1cm); },
]
]
\addlegendimage{ultra thick, sapgreenvar, fill=sapgreenvar}
\addlegendentry{Prompt Length $100$};
\addlegendimage{ultra thick, white, fill=white}
\addlegendentry{Training Step $100k$};
\addlegendimage{ultra thick, white, fill=white}
\addlegendentry{Model Size $1.4b$};
\addlegendimage{ultra thick, lightbluevar, fill=lightbluevar}
\addlegendentry{Prompt Length $300$};
\addlegendimage{ultra thick, white, fill=white}
\addlegendentry{Training Step $140k$};
\addlegendimage{ultra thick, white, fill=white}
\addlegendentry{Model Size $12b$};
\addlegendimage{ultra thick, darkgreenvar, fill=darkgreenvar}
\addlegendentry{Composite All};

\end{axis}
\end{tikzpicture}
}%
\hfill
\begin{tikzpicture}
\begin{axis} [width=0.32\linewidth,height=0.26   \linewidth,yticklabel= {\pgfmathprintnumber\tick\%},ymax=5,ymin=0,ylabel={Extraction Rate (at $l_x=500$)},xlabel={Similarity Threshold ($\delta$)},enlarge x limits=0.05,axis y line*=left,axis x line*=bottom,ylabel near ticks,ylabel style={align=center, text width=4cm, font=\small},xlabel style={align=center, text width=4cm, font=\small},
xtick={0.1, 0.4, 0.7, 1.},ticklabel style = {font=\small},x dir=reverse,
,title style={align=center, text width=6cm, font=\small},
title={\textbf{(c) Comparing Similarity Metrics}}
]

\addplot [name path=all,sapgreenvar,ultra thick] table[x=delta,y=exact] {data/approx_matching_delta.dat};

\addplot [name path=all,chromevar,ultra thick] table[x=delta,y=leven] {data/approx_matching_delta.dat};

\addplot [name path=all,lightvioletvar,ultra thick] table[x=delta,y=lcs] {data/approx_matching_delta.dat};

\addplot [name path=all,darkvioletvar,ultra thick] table[x=delta,y=ngram] {data/approx_matching_delta.dat};

\addplot [name path=all,darkgreenvar,ultra thick] table[x=delta,y=hamming] {data/approx_matching_delta.dat};

\addplot [name path=all,lightbluevar,ultra thick] table[x=delta,y=rougel] {data/approx_matching_delta.dat};


\end{axis}
\end{tikzpicture}%
\raisebox{3em}{
\begin{tikzpicture}[trim left=-3.4em]
\begin{axis}[scale=0.01,
legend cell align={left},
hide axis,
xmin=0, xmax=1,
ymin=0, ymax=1,
legend columns=1,
legend style={font=\scriptsize,/tikz/every even column/.append style={column sep=0.1cm}},
]
]
\addlegendimage{ultra thick, sapgreenvar}
\addlegendentry{Verbatim};
\addlegendimage{ultra thick, chromevar}
\addlegendentry{Levenshtein};
\addlegendimage{ultra thick, lightvioletvar}
\addlegendentry{LCS};
\addlegendimage{ultra thick, darkvioletvar}
\addlegendentry{$n$-gram};
\addlegendimage{ultra thick, darkgreenvar}
\addlegendentry{Hamming};
\addlegendimage{ultra thick, lightbluevar}
\addlegendentry{ROUGE-L};

\end{axis}
\end{tikzpicture}
}%
    \caption{Towards more realistic extraction by combining various churn trends and with approximate matching. \textbf{(a)} Combining two axes at a time, we see a monotonically increasing trend in extraction rates as we gain more points of access to the underlying data, highlighting the growing power of the adversary. \textbf{(b)} Combining all axes of attack results in a significant increase in extraction rate for both standard and deduplicated setups. \textbf{(c)} Various similarity metrics have distinct trends as we decrease the threshold and allow for looser approximations, thus the choice of similarity metric depends on the context of the attack.}
    \label{fig:realistic}
\end{figure*}

\subsection{Combining Multiple Axes of Churn}
\label{sec:realistic_combine}
In the previous section, we saw how churn can impact individual axes of variability. However, a real-world adversary can take advantage of all factors simultaneously, thus significantly increasing their extraction rates. We start by analyzing two axes at a time in Figure \ref{fig:realistic}(a). For all pairs of variability, the overall composite extraction rate (bottom right) is $2-3 \times$ higher than the base setup (top left) and $1.5-2 \times$ higher than the composite extraction rates along one axis (top right and bottom left). Furthermore, when all three axes are combined, depicted in Figure \ref{fig:realistic}(b), the extraction rates grow even higher, albeit with diminishing gains. Thus, a real-world adversary can extract far more training data than shown in existing literature.

\subsection{Computation Cost of Composite Attacks}
\label{sec:cost}

\revision{We now study the computational cost of performing composite extraction attacks. We define our cost in units relative to attacking the Pythia-1.4b model with a prompt length of 100 (the cheapest setup to attack in our multiple axes setting). Cost scales linearly with model size (i.e., attacking the Pythia-2.8b model with prompt length 100 costs 2 units) and quadratically with prompt length (i.e., attacking the Pythia-1.4b model with prompt length 200 costs 4 units), while remaining constant across different checkpoints. The resulting extraction trends for all axes of churn combined under varying cost budgets are shown in Figure \ref{fig:cost}.}

\begin{figure}[h]
    \centering
    \begin{tikzpicture}
\begin{axis} [width=0.3\textwidth,height=0.26   \textwidth,yticklabel= {\pgfmathprintnumber\tick\%},ylabel={Extraction Rate},xlabel={Computation Cost Units},enlarge x limits=0.05,axis y line*=left,axis x line*=bottom,ylabel near ticks,ylabel style={align=center, text width=4cm, font=\small},xlabel style={align=center, text width=4cm, font=\small},
xtick={0, 10, 20, 30, 40, 50},ticklabel style = {font=\small}
]

\addplot [name path=all,chromevar,ultra thick] table[x=cost,y=single] {data/comb_cost.dat};

\addplot [name path=all,darkvioletvar,ultra thick] table[x=cost,y=combgreedy] {data/comb_cost.dat};

\addplot [name path=all,darkgreenvar,ultra thick] table[x=cost,y=comb] {data/comb_cost.dat};

\end{axis}
\end{tikzpicture}%
\raisebox{3em}{
\begin{tikzpicture}[trim left=-5.4em]
\begin{axis}[scale=0.01,
legend cell align={left},
hide axis,
xmin=0, xmax=1,
ymin=0, ymax=1,
legend columns=1,
legend style={font=\scriptsize,/tikz/every even column/.append style={column sep=0.1cm}},
]
]

\addlegendimage{ultra thick, chromevar}
\addlegendentry{Single Greedy};
\addlegendimage{ultra thick, darkvioletvar}
\addlegendentry{Combined Greedy};
\addlegendimage{ultra thick, darkgreenvar}
\addlegendentry{Combined Best};

\end{axis}
\end{tikzpicture}
}
    \caption{\revision{Various strategies of utilizing the adversarial budget and the resulting extraction rates.}}
    \label{fig:cost}
\end{figure}

\revision{We study three different strategies for utilizing the given resources. First, we consider an adversary that greedily attacks the most expensive setup, i.e, prioritizing the largest prompt length, model size, and latest checkpoint, while leaving any leftover resources unused. The extraction rate stays flat until the budget can support a more expensive setup, causing sharp jumps and creating a staircase pattern. This strategy mirrors prior work, where each setup is evaluated in isolation.
Next, we extend the previous strategy to utilize the unused resources, and instead of attacking only a single setup, the adversary now selects multiple setups, still greedily choosing to fit the remaining budget.}

\revision{Finally, we define the third strategy, which is the most effective: searching all combinations to maximize the composite extraction rate, rather than greedily targeting the most vulnerable setups. This consistently outperforms the staircase trend even at the jumps, i.e., even when the greedy strategy utilizes all available budget, showing that combining extractions from less vulnerable setups can exceed the returns of attacking the most expensive setup.}


\subsection{Approximate Matching}
\label{sec:realistic_approximate}
As discussed in \S \ref{sec:combining}, evaluating extraction attacks under verbatim match can underestimate the true risk of extraction. Here, we analyze approximate composite discoverable memorization (Definition \ref{def:approximate_composite_discoverable}) across various similarity metrics $S$ to examine their behaviour under changing $\delta$ in Figure \ref{fig:realistic}(c). Solely for this discussion, we increase the completion length $l_x=500$, to allow for meaningful extraction even with approximate matching.

Our results reveal intriguing trends. First, we study evaluations based on the Levenshtein ratio metric and observe that even the strict threshold of $\delta=0.95$ doubles the extraction attack rate compared to a verbatim match. This threshold signifies a minimum $95\%$ overlap between generated and original text. Clearly, we miss out on a considerable amount of leaked information by relying only on verbatim matches. As $\delta$ decreases, the extraction rate increases exponentially, as the Levenshtein ratio becomes less reliable under looser constraints. We also see similar trends for ROUGE-L scores.

For similarity metrics like longest common substring (LCS), Hamming distance, and $n$-gram matching, even lower values of similarity ($\delta$) can contribute meaningfully to extraction attacks. Unsurprisingly, we observe patterns of increasing extraction rates as before, albeit slower.
The diverse trends underscore the choice of the approximation metric as highly context-dependent. A more thorough examination of which metrics best serve particular applications is left for future work.

\subsection{Data Deduplication}
\label{sec:realistic_deduped}
A commonly recommended solution to memorization is data deduplication, involving the removal of duplicate data entries within a dataset~\citep{carlini2023quantifying}. While costly, data deduplication represents a critical aspect of data curation and has been shown to mitigate extraction risks~\citep{carlini2023quantifying}. To understand the role of data deduplication in our discussion, we repeat our experiments using Pythia models trained on the deduplicated Pile dataset. The results are collected in Figure \ref{fig:realistic}(b).

In line with existing literature, data deduplication reduces the extraction rate. Interestingly, however, we observe persistent trends: the presence of a stronger adversary due to multi-faceted dataset access. 
Thus, while beneficial, data deduplication does not alter our fundamental conclusions; real-world adversaries with multi-faceted access to the underlying data can extract substantial information even post-deduplication. Future work on incorporating more concrete frameworks like differential privacy is needed, to better understand such adversaries, particularly from the perspective of privacy protection under multi-access systems.

\begin{figure*}[t!]
    \centering
    \begin{tikzpicture}
\begin{axis} [width=0.36\linewidth,height=0.25   \linewidth,ymin=0,ylabel={$p$-value ($\downarrow$)},xlabel={Train Dataset Size\vphantom{p}},enlarge x limits=0.05,axis y line*=left,axis x line*=bottom,ylabel near ticks,ylabel style={align=center, text width=4cm, font=\small},xlabel style={align=center, text width=4cm, font=\small},
xtick={500, 600, 700, 800, 900, 1000},ticklabel style = {font=\small},
,title style={align=center, text width=4cm, font=\small},
title={\textbf{(a) Dataset Inference \vphantom{p}}}
]

\addplot [name path=all,sapgreenvar,ultra thick,mark=*,mark options={scale=1,solid}] table[x=datasize,y=single] {data/dataset_inference.dat};

\addplot [name path=all,chromevar,ultra thick,mark=*,mark options={scale=1,solid}] table[x=datasize,y=prompt] {data/dataset_inference.dat};

\addplot [name path=all,darkvioletvar,ultra thick,mark=*,mark options={scale=1,solid}] table[x=datasize,y=size] {data/dataset_inference.dat};

\addplot [name path=all,darkgreenvar,ultra thick,mark=*,mark options={scale=1,solid}] table[x=datasize,y=step] {data/dataset_inference.dat};

\end{axis}
\end{tikzpicture}%
\begin{tikzpicture}
\begin{axis} [width=0.36\linewidth,height=0.25   \linewidth,yticklabel= {\pgfmathprintnumber\tick\%},ylabel={Extraction Rate ($\uparrow$)},xlabel={ROUGE-L Length},enlarge x limits=0.05,axis y line*=left,axis x line*=bottom,ylabel near ticks,ylabel style={align=center, text width=4cm, font=\small},xlabel style={align=center, text width=4cm, font=\small},ticklabel style = {font=\small},
,title style={align=center, text width=4.2cm, font=\small},
title={\textbf{(b) Copyright Infringement}}
]

\addplot [name path=all,sapgreenvar,ultra thick,mark=*,mark options={scale=1,solid}] table[x=length,y=single] {data/copyright.dat};

\addplot [name path=all,chromevar,ultra thick,mark=*,mark options={scale=1,solid}] table[x=length,y=prompt] {data/copyright.dat};

\addplot [name path=all,darkvioletvar,ultra thick,mark=*,mark options={scale=1,solid}] table[x=length,y=size] {data/copyright.dat};

\addplot [name path=all,darkgreenvar,ultra thick,mark=*,mark options={scale=1,solid}] table[x=length,y=step] {data/copyright.dat};

\end{axis}
\end{tikzpicture}%
\raisebox{3em}{
\begin{tikzpicture}
\begin{axis}[scale=0.01,
legend cell align={left},
hide axis,
xmin=0, xmax=1,
ymin=0, ymax=1,
legend columns=1,
legend style={font=\scriptsize,/tikz/every even column/.append style={column sep=0.1cm}},
]
]
\addlegendimage{ultra thick, sapgreenvar}
\addlegendentry{Best Single Setup (Baseline)};
\addlegendimage{ultra thick, chromevar}
\addlegendentry{Composite Prompt Lengths};
\addlegendimage{ultra thick, darkvioletvar}
\addlegendentry{Composite Model Sizes};
\addlegendimage{ultra thick, darkgreenvar}
\addlegendentry{Composite Training Steps};

\end{axis}
\end{tikzpicture}
}
    \caption{\textbf{(a)} $p$-value for dataset inference (lower is better) across different dataset sizes. The results show significant improvement under the composite setup across different prompt lengths. \textbf{(b)} Extraction rate for different ROUGE-L length thresholds, marking potential copyright violations generated by the LLM. Extraction rates with composite setups are consistently higher than the baseline setup.}
    \label{fig:case_studies}
\end{figure*}
\section{Case Studies with Stronger Adversary}
\label{sec:case_studies}

We conclude by highlighting the value of our stronger adversary through various case studies.

\subsection{Detecting Pre-Training Data}
\label{sec:dataset_inference}

Extraction attacks identify whether certain data was included in a model’s training set. This can be valuable in assessing whether a model is trained on proprietary or sensitive data without permission, evaluating data contamination and leakage in various benchmarks, ensuring regulatory compliance with data governance policies, or even academic research to track data contamination.





While membership inference attacks (MIAs) have been used to detect pre-training data, \citet{maini2024llm} argues that in reality, MIAs are as good as random guessing. They show that these attacks learn how to distinguish between \textit{concepts}, and not actual text, highlighting the importance of using IID data of members and non-members to appropriately perform dataset inference.

We borrow their setup and extend it to the composite setting by increasing the size of the training set for learning correlations. Thus, our composite setting can be alternatively seen as an augmentation technique. We record the $p$-value of the null hypothesis \textit{"the dataset was not used for training"} for the Pile dataset in Figure \ref{fig:case_studies}(a), under different sizes of the original training data.

We find that $p$-values for the composite setting across prompt lengths are noticeably lower than the baseline, especially at smaller dataset sizes. Thus, our adversary requires less data to get the same $p$-value. The dataset inference setup by \citet{maini2024llm} requires obtaining IID data, which can be difficult to find. Hence, reducing the amount of such data required can be extremely useful. Interestingly, we did not find strong composition trends across model checkpoints, possibly because membership information can change drastically across models, and thus, combining information from multiple checkpoints might not be helpful.

\subsection{Copyright Infringement}
\label{sec:copyright}

Copyright issues due to LLMs regurgitating their training data have been heavily studied in recent literature. \citet{karamolegkou2023copyright} discusses different thresholds for quoting a text ad verbatim that can be considered a violation of fair use, for example, a $50$ word threshold for magazine articles, chapters, etc., while a $300$ word threshold for books. The authors suggest using ROUGE-L lengths (longest common subsequence) as a measure of text reproduction and potential violations.

We plot the distribution of ROUGE-L lengths for 2,000 randomly chosen examples in Figure \ref{fig:case_studies}(b), comparing the strongest baseline and the composite settings. We find that our adversary produces more potential copyright violations than the baseline, highlighting an underestimation of such risks in existing literature. While copyright law is complex and extraction alone may not imply a violation, our focus is on strengthening the technical underpinnings of copyright issues in LLMs.

\subsection{PIIs Extraction Risk}
\label{sec:pii}

Another commonly studied risk of memorization is extracting personally identifiable information (PIIs). We use the setup of \citet{li2024llm} to create our PII extraction test set from the Pile dataset. We use GLiNER~\citep{zaratiana2024gliner} to detect $2000$ unique PIIs in the Pile dataset, followed by cutting the sentence right before the PII to create the input prompt. These prompts were fed to the model, and the attack is considered successful if the correct PII is generated anywhere within the first 100 tokens, marking the risk of PII leakage~\citep{li2024llm}. 

We record the extraction risk for the best single setup and composite extraction risks across model checkpoints and model sizes. Since the prompts in this setup are of varying lengths, we do not extend our changing prompt lengths setting to this case study. Similar to Definition \ref{def:composite_discoverable}, the composite PII extraction is considered successful if the PII is present in the generation of at least one of the models. The results are collected in Table \ref{tab:pii}, and continuing previous trends, we see a noticeable increase in the extraction rate for an adversary with access to multiple checkpoints.

\begin{table}[h!]
    \centering
    \begin{tabular}{lr}
         \toprule
         \textbf{Setup} & \textbf{Extraction Rate}\\
         \midrule
         Best Single Setup & $22.16\%$ \\
         Composite Model Sizes & $30.97\%$ \\
         Composite Training Steps & $33.07\%$ \\
         \bottomrule 
    \end{tabular}
    \caption{Composite attacks for PII extraction.}
    \label{tab:pii}
\end{table}


\subsection{Closed-Source Models}
\label{sec:closed}

\revision{Our study till now has focused on open-source models due to the availability of training data for verifying the success of our attacks. However, recent work by \citet{decop} has shown how similar studies can be performed even when a model’s training data is undisclosed.}

\revision{\citet{decop} propose DE-COP, a method for detecting memorization in LLMs using a multiple-choice task inspired by counterfactual memorization. The approach presents the target model with four options: one original text passage and three paraphrases generated using a different LLM, in this case, Claude 2. Models tend to choose the original text more frequently if seen during training, thus signaling memorization.}

\revision{The authors validate this method with a novel benchmark - BookTection, which consists of excerpts from 165 books published both before and after 2023. Since closed-source models released in 2022 could not have been trained on books published after 2023, this serves as a reliable non-member set. The books pre-2023 form the member-set. The method compares a model’s performance on each book to a baseline performance (computed from non-member books), using it to assess whether a specific book was part of the model’s training data. Each book can have multiple passages, so the authors combine the accuracy per passage for the detection of the book.}

\revision{For our study, we focus on the experiments on LLaMA-2 70B using the BookTection dataset with varying passage lengths. The authors found that shorter passages yield higher F1 scores: 64, 128, and 256-token passages achieve F1 scores of 0.67, 0.65, and 0.64, respectively. Building on this, we extend their method to a composite setting that combines different lengths. This composite significantly boosts detection performance on LLaMA-2 70B, achieving an F1 score of 0.78,  underscoring the generalizability of our approach.}
\section{Discussion}
\label{sec:discussion}

By highlighting the multi-faceted access adversaries have in the current LLM landscape, our work shows that existing literature greatly underestimates information leakage risks, thus emphasizing the importance of explicitly considering the adversarial perspective and the composability of information leakage in extraction attacks.
Our work provides a foundation for future exploration and defense against more realistic extraction attacks, contributing to a secure and robust management of the risks associated with memorization in LLMs.

\noindent\textbf{Potential Defenses.}
We studied the threats posed by real-world adversaries and showed that existing defence methods (such as data deduplication), while undoubtedly useful, are prone to the same risks of composability. However, we did not propose defense techniques to deal with this adversary.

\revision{Firstly, it's important to recognize that not all instances of discoverable memorization are harmful, and therefore may not require a defense. For instance, defending against improved methods of detecting copyright violations or data contamination is inadvisable. Such efforts could hinder those seeking to determine whether their data was misused by companies or developers.}

\revision{However, when we do want to defend against these attacks in more harmful scenarios, future research could include ways to disrupt multi-faceted access to the dataset. For example, shuffling and re-chunking training sequences for each model can break the link between specific sequences and model behaviour. Since many existing LLM attacks operate at the sequence level~\citep{meeus2024sok}, this simple randomization can significantly increase the difficulty of combining information across checkpoints or model sizes.}

\revision{Other defense strategies, such as anonymizing sensitive data before training~\citep{yuselective} or applying differential privacy (DP) during training~\citep{yudifferentially}, can also help. However, their effectiveness may decline against a stronger adversary. Finally, beyond modifying training, one may wish to defend already trained models against such attacks. While this is challenging, several practical strategies could be potentially helpful. Output perturbation techniques, such as adding noise or rephrasing responses, can reduce information leakage, even though a determined adversary may still bypass them. Access control measures, such as rate limiting and monitoring for suspicious prompts, also offer practical defenses in real-world deployments.}




\noindent\textbf{Beyond Discoverable Memorization.}
Our analysis focuses on the risks posed by extraction attacks under the lens of discoverable memorization. However, our arguments on the increased privacy surface apply to any adversary with multi-faceted access to the underlying data. 
\revisionb{Moreover, our experiments on Pythia and OLMo represent a controlled setup where the underlying models were trained on the exact same data and data order. However, in reality, multiple models from the same family might have some differences in their training.}
Hence, translating our findings to other forms of privacy attacks \revisionb{and dataset homogenization in the real world} is an important direction for future research.

\section*{Acknowledgments}

We thank the anonymous reviewers and the action editor Kai-Wei Chang from TACL, for their continued feedback and comments that helped improve our work. We also thank Pratyush Maini for providing us early access to the dataset inference code~\citep{maini2024llm} before its public release.

Funding support for project activities has been partially provided by Canada CIFAR AI Chair, Google award, MITACS, FRQNT, and NSERC Discovery Grants program.
We also express our gratitude to Compute Canada for their support in providing facilities for our evaluations.

\bibliography{references}
\bibliographystyle{acl_natbib}

\appendix

\section{Unique Extraction Results}
\label{sec:app_unique}

\revision{We provide additional results on the trends of churn across model checkpoints (Figure \ref{fig:unique_steps}) and prompt lengths (Figure \ref{fig:unique_prompts}). We find similar trends as in the main paper.}

\begin{figure*}[t!]
    \centering
    \begin{tikzpicture}
\begin{axis}[width=0.32\linewidth,height=0.24\linewidth,yticklabel= {\pgfmathprintnumber\tick\%},ymin=0,ylabel={Extraction Rate},xlabel={Number of Checkpoints}, enlarge x limits=0.2,xtick=data,ybar=0.5pt,bar width=10pt,axis y line*=left,axis x line*=bottom,symbolic x coords={1, 2, 3, 4, 5, 6, 7, 8, 9},ylabel near ticks,ylabel style={align=center, text width=4cm, font=\small},xlabel style={align=center, text width=4cm, font=\small},
xtick={1, 2, 3, 4, 5, 6, 7, 8, 9},ticklabel style = {font=\small},
,title style={align=center, text width=4.4cm, font=\small},
title={\textbf{(a) Recurrence of Extractions across Training Steps}}
]

\addplot[style={fill=darkgreenvar},draw=none
]
coordinates {(1,0.163)
             (2,0.1)
	 	(3,0.083)
            (4,0.076)
            (5,0.083)
            (6,0.08499999999999999)
            (7,0.1)
            (8,0.148)
            (9,0.8009999999999999)
	     };

\end{axis}
\end{tikzpicture}
\begin{tikzpicture}
\begin{axis}[width=0.32\linewidth,height=0.24\linewidth,yticklabel= {\pgfmathprintnumber\tick\%},ymin=0,xlabel={Training Steps}, enlarge x limits=0.1,xtick=data,ybar=0.5pt,bar width=10pt,axis y line*=left,axis x line*=bottom,symbolic x coords={{100k},{105k},{110k},{115k},{120k},{125k},{130k},{135k},{140k}},ylabel near ticks,ylabel style={align=center, text width=4cm, font=\small},xlabel style={align=center, text width=4cm, font=\small},
xtick={{100k},{110k},{120k},{130k},{140k}},ticklabel style = {font=\small},
,title style={align=center, text width=4cm, font=\small},
title={\textbf{(b) Unique Extractions across Training Steps}}
]

\addplot[style={fill=darkgreenvar},draw=none
]
coordinates {({100k},0.02)
            ({105k},0.012)
            ({110k},0.021)
            ({115k},0.017)
            ({120k},0.009000000000000001)
            ({125k},0.013999999999999999)
            ({130k},0.022000000000000002)
            ({135k},0.021)
            ({140k},0.027)
	     };

\end{axis}
\end{tikzpicture}
\begin{tikzpicture}
\begin{axis}[width=0.32\linewidth,height=0.24\linewidth,yticklabel= {\pgfmathprintnumber\tick\%},ymin=0,xlabel={Training Steps}, enlarge x limits=0.1,xtick=data,ybar=0.5pt,bar width=10pt,axis y line*=left,axis x line*=bottom,symbolic x coords={{100k},{105k},{110k},{115k},{120k},{125k},{130k},{135k},{140k}},ylabel near ticks,ylabel style={align=center, text width=4cm, font=\small},xlabel style={align=center, text width=4cm, font=\small},
xtick={{100k},{110k},{120k},{130k},{140k}},ticklabel style = {font=\small},
,title style={align=center, text width=4cm, font=\small},
title={\textbf{(c) Drop-One Extractions across Training Steps}}
]

\addplot[style={fill=darkgreenvar},draw=none
]
coordinates {({100k},1.619)
            ({105k},1.627)
            ({110k},1.6179999999999999)
            ({115k},1.6219999999999999)
            ({120k},1.63)
            ({125k},1.625)
            ({130k},1.617)
            ({135k},1.6179999999999999)
            ({140k},1.6119999999999999)
	     };

\end{axis}
\end{tikzpicture}
    \caption{\revision{Granular trends of churn across training steps for Pythia. \textbf{(a)} Number of model checkpoints extracting the same information. Even though the majority of samples are extractable by all 9 checkpoints, a significant amount of extractions are unique to a subset, or even a single model. \textbf{(b)} While later model checkpoints contribute more unique extractions, each model, regardless of size, provides significant unique extractions.
    \textbf{(c)} Composite extraction rates after dropping a single model checkpoint (x-axis: dropped checkpoint, y-axis: resulting extraction rate) show that extraction rates remain stable even if we remove the contributions of the last checkpoint.}}
    \label{fig:unique_steps}
\end{figure*}

\begin{figure*}[t!]
    \centering
    \begin{tikzpicture}
\begin{axis}[width=0.32\linewidth,height=0.24\linewidth,yticklabel= {\pgfmathprintnumber\tick\%},ymin=0,ylabel={Extraction Rate},xlabel={Number of Prompts}, enlarge x limits=0.15,xtick=data,ybar=0.5pt,bar width=10pt,axis y line*=left,axis x line*=bottom,symbolic x coords={1, 2, 3, 4, 5, 6, 7, 8, 9, 10},ylabel near ticks,ylabel style={align=center, text width=4cm, font=\small},xlabel style={align=center, text width=4cm, font=\small},
xtick={1, 2, 3, 4, 5, 6, 7, 8, 9, 10},ticklabel style = {font=\small},
,title style={align=center, text width=4.4cm, font=\small},
title={\textbf{(a) Recurrence of Extractions across Prompt Lengths}}
]

\addplot[style={fill=darkgreenvar},draw=none
]
coordinates {(1,0.211)
             (2,0.17500000000000002)
	 	(3,0.145)
            (4,0.153)
            (5,0.214)
            (6,0.22999999999999998)
            (7,0.295)
            (8,0.402)
            (9,0.5660000000000001)
            (10,1.123)
	     };

\end{axis}
\end{tikzpicture}
\begin{tikzpicture}
\begin{axis}[width=0.32\linewidth,height=0.24\linewidth,yticklabel= {\pgfmathprintnumber\tick\%},ymin=0,xlabel={Prompt Lengths}, enlarge x limits=0.1,xtick=data,ybar=0.5pt,bar width=10pt,axis y line*=left,axis x line*=bottom,symbolic x coords={{50},{100},{150},{200},{250},{300},{350},{400},{450},{500}},ylabel near ticks,ylabel style={align=center, text width=4cm, font=\small},xlabel style={align=center, text width=4cm, font=\small},
xtick={{50},{200},{350},{500}},ticklabel style = {font=\small},
,title style={align=center, text width=4cm, font=\small},
title={\textbf{(b) Unique Extractions across Prompt Lengths}}
]

\addplot[style={fill=darkgreenvar},draw=none
]
coordinates {({50},0.016)
             ({100},0.018000000000000002)
	 	({150},0.018000000000000002)
             ({200},0.014)
	 	({250},0.011000000000000001)
             ({300},0.006)
	 	({350},0.015)
             ({400},0.013999999999999999)
	 	({450},0.02)
             ({500},0.08)
	     };

\end{axis}
\end{tikzpicture}
\begin{tikzpicture}
\begin{axis}[width=0.32\linewidth,height=0.24\linewidth,yticklabel= {\pgfmathprintnumber\tick\%},ymin=0,xlabel={Prompt Lengths}, enlarge x limits=0.1,xtick=data,ybar=0.5pt,bar width=10pt,axis y line*=left,axis x line*=bottom,symbolic x coords={{50},{100},{150},{200},{250},{300},{350},{400},{450},{500}},ylabel near ticks,ylabel style={align=center, text width=4cm, font=\small},xlabel style={align=center, text width=4cm, font=\small},
xtick={{50},{200},{350},{500}},ticklabel style = {font=\small},
,title style={align=center, text width=4cm, font=\small},
title={\textbf{(c) Drop-One Extractions across Prompt Lengths}}
]

\addplot[style={fill=darkgreenvar},draw=none
]
coordinates {({50},3.4979999999999998)
             ({100},3.496)
	 	({150},3.496)
             ({200},3.501)
	 	({250},3.5029999999999997)
             ({300},3.508)
	 	({350},3.499)
             ({400},3.5000000000000004)
	 	({450},3.4939999999999998)
             ({500},3.434)
	     };

\end{axis}
\end{tikzpicture}
    \caption{\revision{Granular trends of churn across prompt lengths for Pythia. \textbf{(a)} Number of prompts extracting the same information. Even though the majority of samples are extractable by all 10 prompt lengths, a significant amount of extractions are unique to a subset, or even a single prompt length. \textbf{(b)} While longer prompt lengths contribute more unique extractions, each prompt length provides some unique extractions.
    \textbf{(c)} Composite extraction rates after dropping a single prompt length (x-axis: dropped prompt length, y-axis: resulting extraction rate) show that extraction rates remain stable even if we remove the contributions of the longest prompt length.}}
    \label{fig:unique_prompts}
\end{figure*}

\end{document}